\documentclass[english,a4paper]{article}
\usepackage[T1]{fontenc}
\usepackage[cp1250]{inputenc}
\usepackage{amssymb}
\usepackage{hyperref}
\usepackage{amsfonts}
\usepackage{graphicx}

\def\openone{\leavevmode\hbox{\small1\kern-3.3pt\normalsize1}}
\begin{document}

\title{Comparative study of monotonically convergent optimization
  algorithms for the control of molecular rotation}

\author{M. Ndong, M. Lapert\footnote{Institut
f\"ur Quanteninformationsverarbeitung, Universit\"at Ulm, D-89069
Ulm, GERMANY}, C. Koch\footnote{Theoretische Physik, Universit\"at Kassel,
  Heinrich-Plett-Str. 40, 34132 Kassel,Germany} and D. Sugny\footnote{Laboratoire Interdisciplinaire
Carnot de Bourgogne (ICB), UMR 5209 CNRS-Universit\'e de
Bourgogne, 9 Av. A. Savary, BP 47 870, F-21078 DIJON Cedex,
FRANCE, dominique.sugny@u-bourgogne.fr}}

\maketitle

\begin{abstract}
We apply two different monotonically convergent optimization
algorithms to the control of molecular rotational dynamics by laser
pulses. This example represents a quantum control problem where the
interaction of the system with the external field is non-linear.
We test the validity and accuracy of the two methods on the key
control targets of producing molecular
orientation and planar delocalization at zero temperature,  and maximizing permanent
alignment at non-zero temperature.
\end{abstract}


\section{Introduction}
\label{sec:intro}
Optimal control theory is nowadays a mature mathematical discipline
with a wide range of applications in science and engineering
\cite{pont}. The technique has been used with success in quantum
mechanics since the beginning of the 1990s
\cite{rice,shapiro,tannorbook,brif} to control spins, atoms
and molecules by external electromagnetic fields. Control problems can
be tackled by two different types of approaches, geometric
\cite{khaneja,lapertglaser,simutime} and  numerical methods
  \cite{kosloff,somloi,zhu,grapeino,maday,ohtsukimono,gross}
for quantum systems of
low and high dimension, respectively. It is this second aspect which
is at the core of this article. Numerical optimal control
algorithms can roughly be divided into Gradient ascent algorithms
\cite{grapeino} and Krotov's method
\cite{somloi, KonnovAC99, SklarzPRA02, JosePRA03, KrotovDok08, KrotovAC09, reich}.
The latter guarantees monotonic convergence independent of the
specific choice of optimization functional, type of interaction
between system and external control, and equations of motion. In the
quantum control literature, Krotov's method was first established for
dipole transitions, where the interaction of the system with the control
field is linear~\cite{somloi,JosePRA03,koch04,JosePRA08}. In
recent years, several modifications to the known algorithms
have been brought forward  to account for the non-linear case, a
problem which arises
naturally in a variety of control problems in atomic and molecular
physics. In particular it occurs when the intensity of the laser
field is sufficiently large, so that the linear model is no longer a
good approximation of the dynamical system. While the generalization
is straightforward for gradient algorithms, the extension of
the monotonic approach is more
involved~\cite{ohtsukinonlinear,lapertalgo,reich}. Here, our goal
is  to explore the efficiency of two different schemes
of monotonically convergent optimization algorithms for the control of
a molecule interacting non-linearly with the control
field.

The control of molecular rotation 
\cite{friedrich,seideman,stapelfeldt,averbukh,salomon,liao,viellard,lapertthz}, for which
such non-linear models are well established \cite{sakai,tehini}, is
used as a testbed case to analyze the features of these
algorithms. A first modification of a monotonically convergent
algorithm to account for a non-linear interaction with the control
assumes the cost to be quadratic in the field and decomposes the
control  into $n$ components for a nonlinearity of order
$n$~\cite{ohtsukinonlinear}. The decomposition leads to $2n$
Schr\"odinger equations that need to be solved, $n$ for the wave
function and $n$ for the
adjoint state. This can be numerically costly. The approach was
successfully applied to the control of molecular orientation and
alignment \cite{ohtsukialign}. At the same time, some of us proposed
a new algorithm using only one component of the wave function
\cite{lapertalgo}. This comes at the price of changing the cost
functional. Instead of penalizing the intensity of the field,
i.e., the square of the control parameter, it penalizes a higher
exponent, the value of which depends on the order of the
non-linearity. The algorithms of
Refs.~\cite{ohtsukinonlinear,lapertalgo} have recently
been compared \cite{compar_ohtsuki}. In the case of a two-color control strategy for molecular orientation, it was shown that the efficiency of the two optimized solutions designed by the two algorithms was similar.
In parallel, it has been mentioned that the Krotov method allows for
constructing a monotonically convergent algorithm with the standard cost
functional penalizing the field intensity \cite{reich}. Here, we
examine this claim and perform an extensive comparison with the
Lapert algorithm \cite{lapertalgo}.
We analyze the efficiency, numerical cost and  structure of the
optimized solutions obtained by the two approaches. The rotational
dynamics of a diatomic
molecule driven by an electromagnetic field will be used as an
illustrative example.

The remainder of this paper is organized as follows. The molecular
model  is presented in Sec. \ref{sec:model}. Section \ref{sec:appli} is
devoted to the application of the two algorithms to the control
objectives of controlling  molecular orientation and planar delocalization at zero
temperature, and producing permanent alignment at non-zero
temperature. We conclude in Sec.~\ref{sec:conclusions}.
Appendix~\ref{app}
summarizes briefly the two optimization algorithms.

\section{The model}
\label{sec:model}
We consider the control of the rotational dynamics of the linear CO
molecule described in a rigid rotor approximation and driven by
the electric field $\vec{E}(t)$. The field is expressed as follows:
\begin{eqnarray}
  \label{eq:field-ana}
  \vec{E}(t) &=&
  \epsilon_x(t)\cos(\omega t+\Phi_x)\vec{e}_x+
  \epsilon_z(t)\cos(\omega t+\Phi_y)\vec{e}_y+
  \nonumber  \\
  &&
  \epsilon_z(t)\cos(\omega t+\Phi_z)\vec{e}_z,
\end{eqnarray}
where  $\epsilon_\nu(t)$, $\vec{e}_\nu$, and $\Phi_\nu$, $\nu=x, \,y,\, z$ are
the amplitude, the unit vector and the phase along the
$\nu$-axis, respectively.
The Hamiltonian of the system can be written as \cite{sakai,tehini}:
\begin{equation}
  \label{eq:ham1}
 H = BJ^2 + \vec{\mu}\cdot\vec{E}(t) +
  \bold{\alpha}\cdot\vec{E}^2(t) +\bold{\beta}\cdot\vec{E}^3(t),
\end{equation}
where $B$ is the rotational constant. The first term of the right-hand
side of Eq. (\ref{eq:ham1}) is the
field-free rigid-rotor Hamiltonian. Its eigenstates are the spherical harmonics
denoted by $|j,m\rangle$, with $j\ge 0$ and $|m|\leq j$. The
operators $\vec{\mu}$, $\bold{\alpha}$ and $\bold{\beta}$ are associated,
respectively, to the permanent dipole moment and the polarizability
and hyperpolarizability tensors.
The spatial position of the diatomic molecule is given in
the laboratory frame by the spherical coordinates $(\theta,\phi)$.

We first study the interaction of the molecule with a  non-resonant
laser field, polarized linearly along the $z$-axis of the laboratory
frame. In this case,
the variable $\theta$ is the angle between the molecular axis and the
polarization vector of the electric field. The
Hamiltonian~(\ref{eq:ham1})  then simplifies to
\begin{eqnarray}
  \label{eq:hamZ}
  H &=& B J^2 -
  \mu_0
  \cos\theta E_z(t)
  - \frac{1}{2}\left[
    \Delta\alpha
  \cos^2\theta +\alpha_\perp\openone
 \right ]E_z^2(t)
  \nonumber  \\
  &&
  - \frac{1}{6}\left[
    (\beta_\parallel-3\beta_\perp)
    \cos^3\theta +3\beta_\perp\cos\theta
 \right ]E_z^3(t),
\end{eqnarray}
where $\Delta\alpha = \alpha_\parallel-\alpha_\perp$. The parameter
$\mu_0$ is the permanent dipole moment and the coefficients
$\alpha_\parallel$, $\alpha_\perp$, $\beta_\parallel$ and
$\beta_\perp$ denote, respectively, the polarizability and
hyperpolarizability components of the molecule with the labels
$\parallel$ and $\perp$ indicating the components parallel and
perpendicular to the
internuclear axis. The numerical values used in our simulations for
the different molecular parameters are reported in
Tab. \ref{tab:param}. For details see Ref. \cite{lapertalgo}.
\begin{table}[tb]
  \centering
  \begin{tabular}{|c|c|c|c|c|c|}\hline
    $B$  (cm$^{-1}$)& $\mu_0$ (a.u.) & $\alpha_\parallel$ (a.u.) &
    $\alpha_\perp$ (a.u.)&  $\beta_\parallel$ (a.u.) & $\beta_\perp$ (a.u.) \\ \hline
    1.9312   & 0.112   &  15.65 &   11.73   & 28.35  & 6.64 \\
    \hline
  \end{tabular}
\caption{Numerical values of the different molecular parameters.}
  \label{tab:param}
\end{table}
If we further assume that the frequency $\omega$ of the laser field is
much higher than the rotational frequencies and non-resonant with
respect to all rovibronic transitions, we can average over the fast
oscillations of the electric field in Eq. (\ref{eq:hamZ}) and obtain
\cite{tehini,sakai}:
\begin{eqnarray}
  \label{eq:hamZ_env}
  H(t) &=& BJ^2
  - \frac{1}{4}\left[
    \Delta\alpha
  \cos^2\theta +\alpha_\perp\openone
 \right ]\epsilon_z^2(t)
  \nonumber  \\
  &&
  - \frac{1}{8}\left[
    (\beta_\parallel-3\beta_\perp)
    \cos^3\theta+3\beta_\perp\cos\theta
 \right ]\epsilon_z^3(t).
\end{eqnarray}

As a second example we consider the
interaction of the CO molecule with a pulse that is elliptically polarized in
the $(x,y)$-plane. We neglect here the hyperpolarizability
term of the interaction since it does not play a quantitative role in this case.  
After optical-cycle averaging as above, the
corresponding Hamiltonian is expressed as:
\begin{eqnarray}
  \label{eq:hamXY}
  H &=& BJ^2
  - \frac{1}{4}\left[
    \left(\Delta\alpha
  \cos^2\theta_x +\alpha_\perp \openone\right) \epsilon_x^2(t)+
    \right. \nonumber \\
    && \left.
    \left(\Delta\alpha
  \cos^2\theta_y+\alpha_\perp \openone\right) \epsilon_y^2(t)+
    \right. \nonumber \\
    && \left.
    2\Delta\alpha
    \cos(\Phi_x-\Phi_y)\cos\theta_x \cos\theta_y \epsilon_x(t)\epsilon_y(t)
 \right ],
\end{eqnarray}
where $\cos\theta_x = \sin\theta\cos\varphi$ and
$\cos\theta_y = \sin\theta\sin\varphi$.

In the case of zero rotational temperature ($T=0$ K),
the time evolution of the system is described by the time-dependent
Schr\"odinger equation,
\begin{equation}
  \label{eq:evolwp}
    i\frac{\partial}{\partial t} |\psi(t)\rangle = H(t)|\psi(t)\rangle,
\end{equation}
where $|\psi(t)\rangle$ is the wave function of the system at
time $t$. The Liouville equation
is used to describe the time evolution for $T \neq 0$ K:
\begin{equation}
  \label{eq:evolrho}
  i\frac{\partial}{\partial t} \rho(t) = [H(t),\rho(t)],
\end{equation}
where $\rho(t)$ represents the density matrix associated with the
system at time $t$.
Equations (\ref{eq:evolwp}) and (\ref{eq:evolrho}) are solved
numerically with the split operator algorithm \cite{Feit82:412}.
The Hamiltonian is represented in spherical harmonics $|j,m\rangle$ where
all matrix elements are known analytically.
We use atomic units unless otherwise specified.


\section{Numerical results}
\label{sec:appli}
We explore three different control targets presenting a comparative
study of the Krotov and Lapert algorithms. The
technical details of the algorithms are briefly reviewed in Appendix
\ref{app}.

\subsection{Orientation dynamics driven by a linearly polarized field}
\label{subsec:orient}
\begin{figure}[tb]
  \centering
  \includegraphics[width=0.98\linewidth]{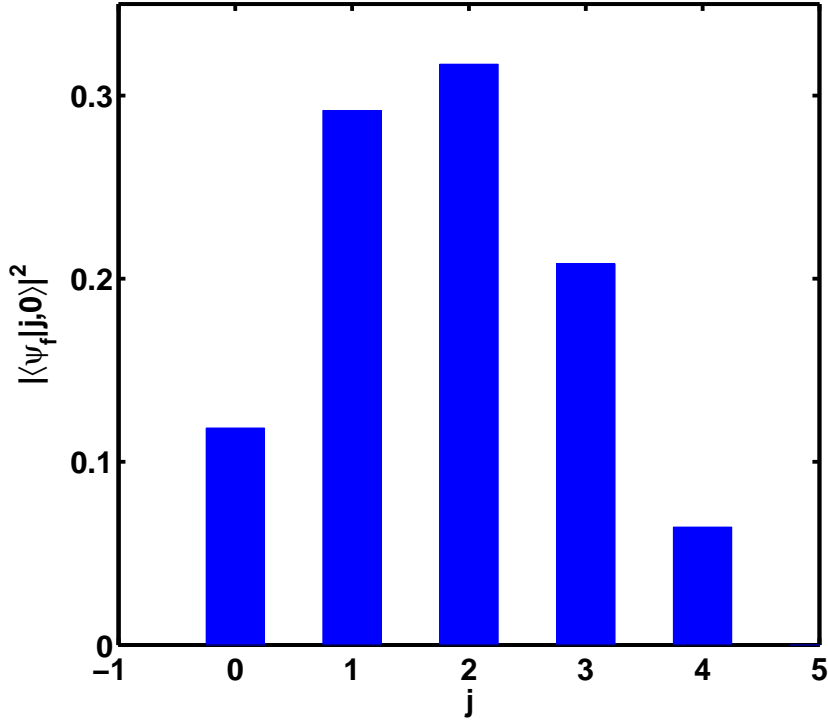}
  \caption{(color online)
    Population of the target state maximizing orientation of the CO molecule
    along the $z$-axis. The corresponding wave function is given by
    $|\psi_f\rangle  \approx
    0.34|0,0\rangle +
    0.54|1,0\rangle +
    0.56|2,0\rangle +
    0.46|3,0\rangle +
    0.25|4,0\rangle
    $
  }
\label{fig:target}
\end{figure}

We first investigate the control of molecular
orientation by a  field linearly polarized along the $z$-axis. In this
case, the dynamics is described by Eq. (\ref{eq:hamZ}). The
control duration $t_f$ is chosen to be equal to one rotational period
$T_{per}$ of the molecule, $T_{\mathrm{per}} \approx 8.6$ ps. We
consider a finite Hilbert space of size $j_{max}=15$, which is
sufficient for the intensity of the laser field used here. The
expectation value
$\langle\cos\theta\rangle$ is taken as a quantitative measure of the
orientation. The molecule is oriented
when $|\langle\cos\theta\rangle|\simeq 1$. Following
Refs. \cite{sugnywave,sugnydensity}, we do not maximize this
expectation value but a
target state $|\psi_f\rangle$ maximizing $|\langle\cos\theta\rangle |$
in a sub-Hilbert space of finite dimension defined by $j\leq j_f$. The
details of $|\psi_f\rangle$ can be found in Refs. \cite{sugnywave, sugnydensity}. Figure \ref{fig:target} shows
the projection of the target state onto the
 eigenstates $|j,0\rangle$ of the molecule, with $j_f=4$.
A Gaussian pulse of 144 fs full width at half maximum (FWHM), centered
at $t_0 = T_{per}/5$ is taken as guess field for all optimizations
discussed in this section:
\begin{equation}
  E(t) = E_0 e^{\frac{(t-t_0)^2}{2\sigma^2}},
\end{equation}
where $\frac{1}{2}\epsilon_0 c E_0^2 = 10^{12}$ W/cm$^2$
is the peak intensity of the laser field and the parameter $\sigma$ is
defined such that
FWHM = $2\sqrt{2\ln 2} \sigma$. This choice of guess field is standard in the control of molecular orientation \cite{sugnywave}. It provides an efficient initial solution with a population transfer from the state $|0,0\rangle$ to a superposition of $|j,0\rangle$ states with $j=0,1,\cdots,j_f$. The cost functional is defined by:
\begin{equation}\label{cost}
C=|\langle \psi_f|\psi(t_f)\rangle |^2-\lambda \int_0^{t_f}(E(t)-E_{ref}(t))^n/S(t)dt,
\end{equation}
where $n=2$ and $n=4$ for the Krotov and Lapert algorithms,
respectively. The  parameter $\lambda$
penalizes the pulse energy and $E_{ref}(t)$ is a pulse reference. The
function $S$, given by $S(t)=\sin^2(\pi t/t_f)$, suppresses pulse
amplitude  at the beginning
and end of the time window, ensuring a pulse that is smoothly switched
on and off. We denote the final fidelity of the control
by $C_{t_f}=|\langle \psi_f|\psi(t_f)\rangle |^2$.

\begin{figure}[tb]
  \centering
  \includegraphics[width=0.98\linewidth]{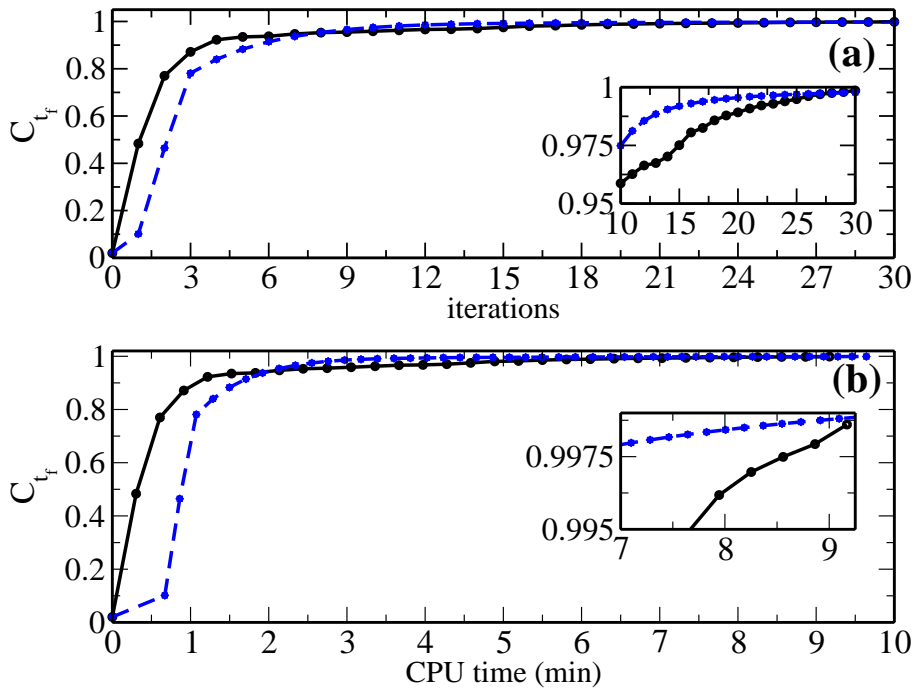}
  \caption{(color online)
    Convergence of the Lapert and Krotov algorithms measured
    by the final cost $C_{t_f}$ plotted as a function of
    the number of iterations (top panel) and  $C_{t_f}$
    plotted as a function of the CPU time (bottom
    panel).
    The results for  the Lapert formulation are depicted in black solid line with circles,
  while a blue dashed line with stars represents the results with the Krotov algorithm.}
  \label{fig:JT_scheme1}
\end{figure}

We first analyze the role of the parameter $\lambda$ in the two
algorithms.  The results reported in Tab. \ref{tab:conver_lambda_a}
show that the Krotov algorithm requires smaller $\lambda$ values to
converge to a high fidelity $C_{t_f}$ than the Lapert method.
\begin{table}[tb]
  \centering
  \begin{tabular}{|c|c|c|c|c|c|}\hline
    $\lambda$  & $C_{t_f}^{\mathrm{lp}}$  & $C_{t_f}^{\mathrm{kr}}$  &
    $E_{\max}^{\mathrm{lp}}$ &  $E_{\max}^{\mathrm{kr}}$   \\ \hline
      5\,10$^{6}$   & 0.9892   &  0.0205 &   5.5\,10$^{-3}$   & 5.5\,10$^{-3}$  \\
      5\,10$^{4}$   & 0.9993   &  0.0205 &   5.5\,10$^{-3}$   & 10$^{-2}$  \\
      5\,10$^{2}$   & 0.9996   &  0.0205 &   5.5\,10$^{-3}$   & 2.8\,10$^{-2}$  \\
      5\,10$^{-1}$   &  $--$   &  0.5789 &   $--$             & 5.5\,10$^{-3}$  \\
      5\,10$^{-2}$   &  $--$   &  0.9959 &   $--$             & 5.5\,10$^{-3}$  \\
      5\,10$^{-3}$   &  $--$   &  0.9944 &   $--$             & 5.5\,10$^{-3}$  \\
    \hline
  \end{tabular}
\caption{Analysis of the convergence of the two algorithms with
  respect to the parameter $\lambda$.
  $C_{t_f}^{\mathrm{lp}}$  and  $C_{t_f}^{\mathrm{kr}}$ are  the fidelities
  obtained from Krotov and Lapert algorithms, respectively. $E_{\max}^{\mathrm{lp}}$ and $E_{\max}^{\mathrm{kr}}$
correspond to the maximum amplitude in absolute value of the two optimized solutions. The number of
  iterations is set to 20. In the Lapert algorithm, small values of $\lambda$ lead to very fast oscillations of the corresponding optimized field, which are not physically relevant. Therefore results are not indicated in
columns 2 and 4 when $\lambda$ is smaller than $5\times 10^{2}$.}
  \label{tab:conver_lambda_a}
\end{table}
In order to observe a convergence with realistic optimized
pulses, the parameter $\lambda$ should be larger than 10$^2$
for the Lapert method and  lower than 0.1 for the Krotov one. This
difference is easily understood from the fact that, in the Krotov formulation, the
running cost is a quadratic function of the electric field,
cf. Eq.~(\ref{cost}), while in the algorithm by Lapert et al., this
power is 4. Since in atomic units, $|E(t)-E_{\mathrm{ref}}|<1$,
the same order of magnitude for the two running costs is obtained for
$\lambda^{\mathrm{Lp}} \ge 10^2 \lambda^{\mathrm{Kr}}$. Note that the parameter $\lambda^{\mathrm{Kr}}$ has
to satisfy the relation (\ref{eq:cond-delta2}) in order for the Krotov algorithm to be monotonic.
The fact that there is no constraint in the choice of $\lambda^{\mathrm{Lp}}$ allows more flexibility
in the use of the Lapert algorithm. However, one should keep in mind that
large values of $\lambda^{\mathrm{Lp}}$ are required
 in order to avoid fast oscillations in the optimized solution.
In this section, the two coefficients will be
fixed at $\lambda^{\mathrm{Lp}} =  5\times 10^{6}$  and
$ \lambda^{\mathrm{Kr}} = 5\times 10^{-2}$.
The evolution of the final fidelity $C_{t_f}$
as a function of the number of iterations and the CPU time is displayed in
Fig.~\ref{fig:JT_scheme1}. While the Lapert formulation converges faster initially, the insets of
  Fig.~\ref{fig:JT_scheme1} show that the Krotov algorithm becomes faster
when the fidelity is close to 1.

The Lapert algorithm is however more costly in terms of computer time (CPU time).
The faster convergence of Krotov's method
in terms of CPU time is not surprising since
in the Lapert formulation, at each iteration,
the roots of a polynomial of power 3 need to be determined in order to compute
the updated new pulse, see the appendix \ref{subsec:lapert} for details.
Figure \ref{fig:field_scheme1} compares the optimized pulses obtained from the Lapert
algorithm, Fig.~\ref{fig:field_scheme1}a and from the Krotov one,
Fig.~\ref{fig:field_scheme1}c. Figure \ref{fig:field_scheme1}b displays
the guess field considered for the two algorithms. Note that the structure of the Krotov solution
is very simple, since the optimized field is mainly composed of the guess pulse plus additional small deformations.
The solution designed by the Lapert algorithm is rather more complex, in the sense that fast
oscillations appear between the middle and the end of the optimization time interval. As already pointed out in \cite{lapertalgo}, this  behavior seems to be quite general with a cost functional penalizing the power 4 of the field. Note that spectral filters can be added to avoid such oscillatory structures \cite{viviespectrum,lapertspectrum}.
\begin{figure}[tb]
  \centering
  \includegraphics[width=0.98\linewidth]{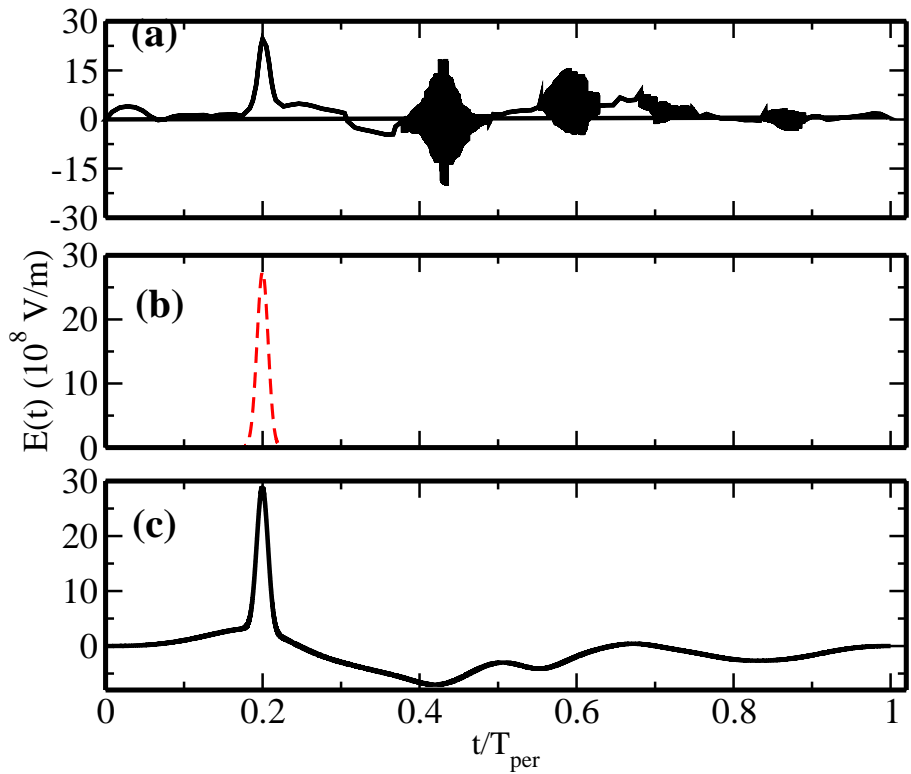}
  \caption{(color online)
    Comparison of the optimized pulses obtained from the  Krotov (c) and Lapert (a) algorithms.
  The panel (b) depicts the initial field used for the two algorithms.}
  \label{fig:field_scheme1}
\end{figure}

\begin{figure}[tb]
  \centering
  \includegraphics[width=1.1\linewidth]{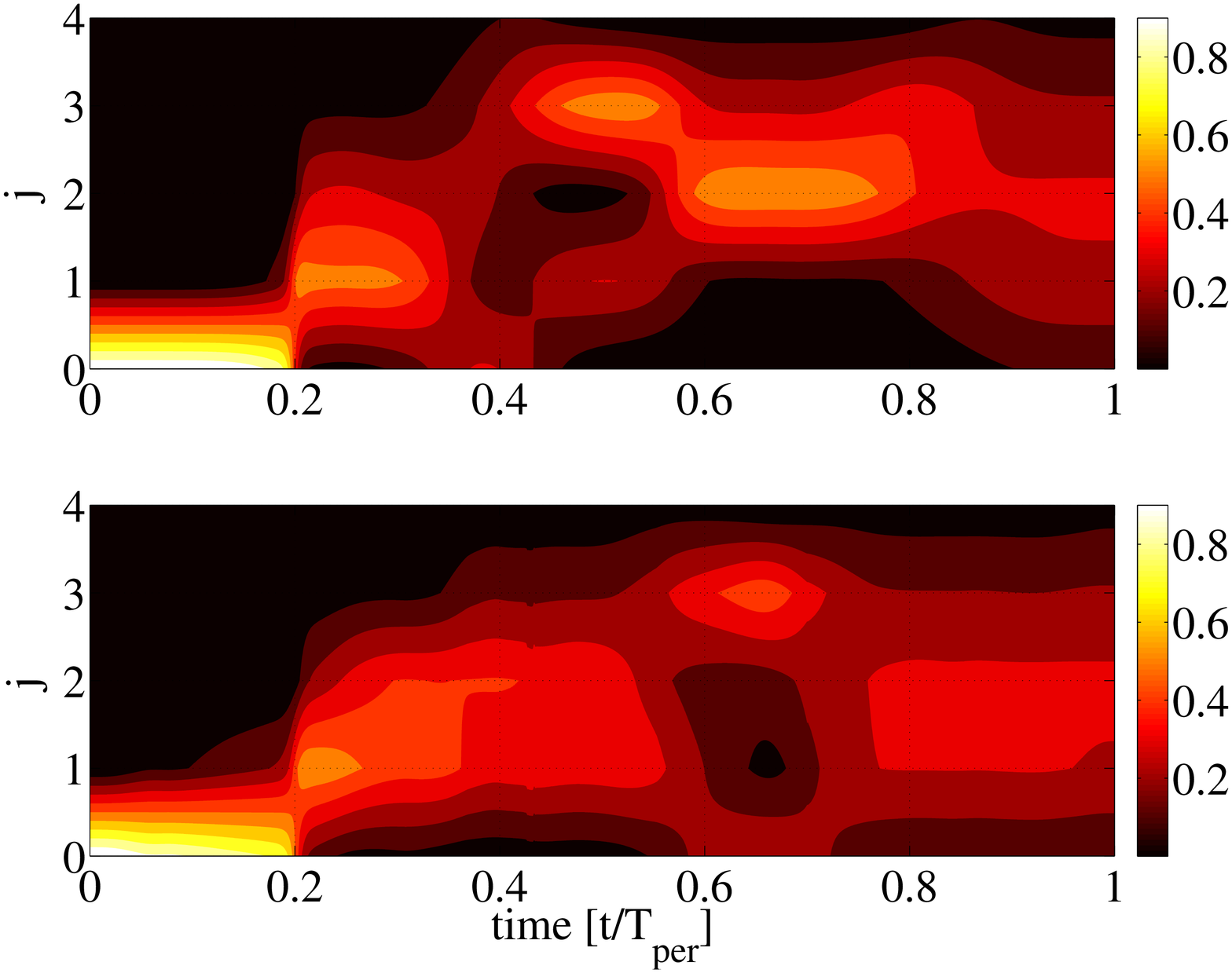}
  \caption{(color online)
    Projection of the time-dependent wave function onto the eigenstates
    of the system. The top and bottom panels represent respectively the dynamics induced by the pulses
    optimized with the Krotov and Lapert algorithms.}
  \label{fig:dyn}
\end{figure}

The dynamics induced by the two optimized pulses is plotted in Fig. \ref{fig:dyn}.
More precisely, figure \ref{fig:dyn} displays the projection
of the wave function onto the molecular eigenstates as a function of
time. The two dynamics show similar features in the first fifth
of the optimization time interval, $[0,0.2\times t_f]$. Most of the population remains in
the ground state, $|j=0,m=0\rangle$, since,
in this time interval, both optimized pulses are almost zero. At time
$t=0.2\times t_f$, both optimized solutions contain a kick which leads to a
superposition of states with $j=0,\ldots, 3$.
For the dynamics induced by the Krotov optimized pulse, we observe that
most of the population is concentrated in states $|j=3,0\rangle$ and
$|j=4,0\rangle$ during the time interval $[0.4\times t_f,0.6\times t_f]$, while
these states are populated for $t\in [0.6\times t_f,0.8\times t_f]$ in the Lapert
case.
In the time interval $[0.6\times t_f,0.8\times t_f]$, the dynamics
induced by the Krotov optimized field shows a  superposition  of
states $|1,0\rangle$, $|2,0\rangle$ and $|3,0\rangle$, more or less
similar to the target state.
In the final step, i.e., in the time interval $[0.8\times t_f,t_f]$,
the small oscillations of the Lapert and Krotov fields are responsible
for the complete transfer of the superpositions to the target
state.
\subsection{Delocalization in the ($x,y$)- plane}
\label{subsec:deloc}
The second example is dedicated to  controlling the orientation of
the angular momentum of the CO molecule along the $z$-axis of the
laboratory frame. The degree of orientation is given here by $\langle J_z\rangle /\sqrt{ \langle J^2\rangle}$,
where $J_z$ is the $z$- component of the angular momentum $J$. This aspect has been recently studied in a
series of works, both theoretically
\cite{fleicher,sugnyplanar2,cogwheel} and experimentally
\cite{kitano,sugnyplanar1,milner}. In particular, it has been shown in
Ref. \cite{cogwheel} that orientation of the angular momentum can
be achieved by a sequence of two short laser pulses, properly
delayed and polarized at 45 degrees with respect to each other. Here
we revisit this control problem using the two
monotonically convergent algorithms. An elliptical polarization is
considered to realize this orientation.
The corresponding Hamiltonian is given by Eq.~(\ref{eq:hamXY}).
Since the angular momentum of a diatomic molecule is classically
orthogonal to the molecular axis, its alignment along the laboratory
$z$- axis is equivalent to a delocalization of the molecular axis in the ($x,y$)- plane.
This delocalization can also be interpreted as a minimization of the expectation
value $\langle\cos^2\theta\rangle$ \cite{sugnyplanar2}.

Let us consider a state of the form $|j,\pm j\rangle$. Straightforward
computation shows that
\begin{equation}
  \label{eq:cos2jj}
  \langle j,\pm j| \cos^2\theta |j,\pm j\rangle = \frac{1}{2j+3}.
\end{equation}
When $j \rightarrow \infty$, the right hand side of Eq.~(\ref{eq:cos2jj})
converges to its minimum value, 0. Therefore,  in a sub-Hilbert space
of finite dimension, the states $|j,\pm j\rangle$
minimize $\langle\cos^2\theta\rangle$ for large $j$. In other words,
these states maximize the  delocalization of the molecular axis in the
($x,y$)-plane. Consequently, the states $|j, j\rangle$ and the states
$|j,-j\rangle$
maximize and minimize $\langle J_z \rangle$, respectively \cite{cogwheel}.

At $T=0$ K, the initial state is $|0,0\rangle$ and
$|4,4\rangle$ is taken as the target state.
The minimum expectation value that can be
reached with this choice is of the order of $ \langle 4,\pm 4|
\cos^2\theta |4,\pm 4\rangle \approx 0.1$. Note that the more we
increase $j$, the better the delocalization becomes. However, both
difficulty and numerical cost  increase with $j$. Therefore, target states of the form
$|j,\pm j\rangle$ with $j>4$ will not be analyzed.

\begin{figure}[tb]
  \centering
  \includegraphics[width=0.98\linewidth]{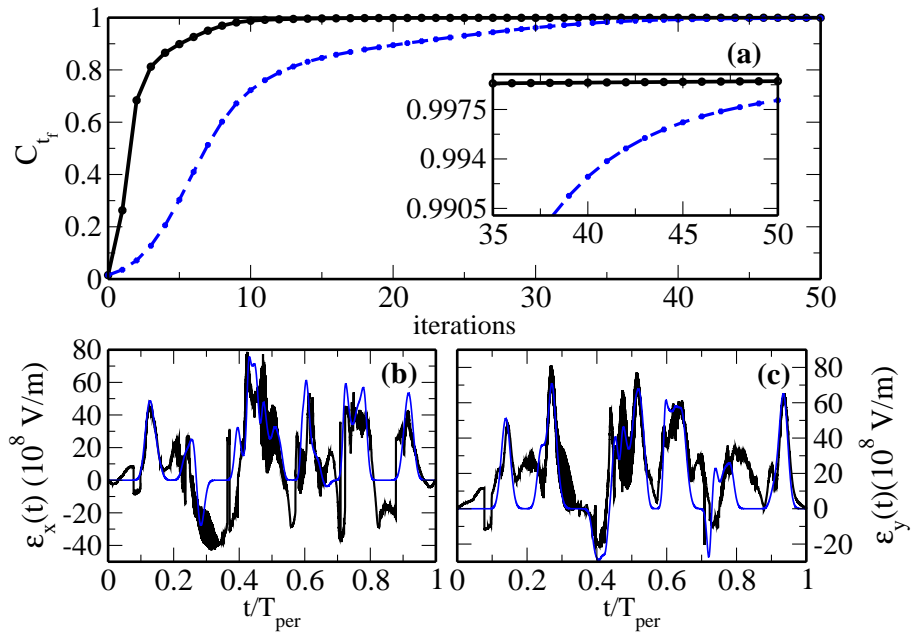}
  \caption{(Color online)
    (a): Convergence of the Lapert and the Krotov algorithms measured
    by the final cost $C_{t_f}$.
     $C_{t_f} = |\langle 4,4|\psi(t_f)\rangle|^2$ (top panel) is
    plotted as a function of the number of iterations.
    (b): $x$- components of the optimized fields. (c): $y$- components the optimized fields.
    The black and
    blue (dark gray) curves depict respectively the results derived
    with the Lapert and the Krotov algorithms.}
  \label{fig:JT_jj}
\end{figure}

When the relative phase $\Phi_x-\Phi_y$ is chosen so that the cross
term of the Hamiltonian (\ref{eq:hamXY}) vanishes, the dynamics cannot distinguish the states $|j,
j\rangle$ and $|j,-j\rangle$ (see \cite{sugnyplanar2} for the analytical proof). Here, in
order to get a completely controllable system, the relative phase $\Phi_x-\Phi_y$ is
set to $\pi/4$. 
The guess field is constructed as a series of Gaussian pulses of 150
fs FWHM for each component, $\epsilon_x$
and $\epsilon_y$. We have chosen $\lambda^{\mathrm{Lp}} =  10^{5}$  and
$ \lambda^{\mathrm{Kr}} = 0.1$. Our choice of a large $\lambda$ value for
 the Lapert algorithm is motivated by the fact that
 small values induce very fast oscillations of the optimized
 control field, which are physically and numerically not very
 interesting. For 50 iterations, the target state
 is reached with a probability of the order of 0.99 for the two algorithms.
The convergence of the cost $C_{t_f}$ (top panel), plotted as a function of the number of iterations,
is shown in Fig.~\ref{fig:JT_jj} for the two algorithms. The corresponding optimized pulses are shown in the bottom panels.
The convergence to the target state is similar to the one observed in the first example. In particular, the Lapert algorithm converges faster initially, the Krotov algorithm remaining more efficient, specifically in terms of CPU time. In addition, Fig. \ref{fig:JT_jj} clearly shows that the structure of the Krotov solution is simpler than the Lapert one.
\subsection{Control of a thermal rotational sample}
\label{subsec:sample}
The last example concerns the control of a thermal sample at non-zero
temperature. The initial state is given by the Boltzmann distribution,
which can be written as follows:
\begin{equation}
  \label{eq:rho_0}
  \rho_0 = \frac{1}{Z} \sum_j \sum_{m=-j}^j e^{-\frac{Bj(j+1)}{k_BT}}
  |j,m\rangle \langle j,m|\,,
\end{equation}
where $T$ is the temperature, $k_B$ the Boltzmann constant and $Z$ the
partition function, which is expressed as
\begin{equation}
  \label{eq:ZZ}
  Z = \sum_j\sum_{m=-j}^j e^{-\frac{Bj(j+1)}{k_BT}}.
\end{equation}
The control aims at reaching the state that maximizes the permanent
alignment of the molecule along the $z$-axis. The molecular alignment
is measured by the expectation value $\langle \cos^2\theta\rangle$,
which can be written as the sum of two terms:
\begin{equation}
\langle\cos^2\theta\rangle =\langle\cos^2\theta\rangle _p+\langle \cos^2\theta\rangle _c,
\end{equation}
where $\langle\cos^2\theta\rangle
_p=\sum_{j,m,m'}\rho_{jm,jm'}C_{jm,jm'}$ and $\langle \cos^2\theta
\rangle _c =\sum_{j\neq j',m,m'}\rho_{jm,j'm'}C_{jm,j'm'}$. The
coefficients $C_{jm,j'm'}$  denote the matrix
elements of the operator $\cos^2\theta$.
Partitioning the alignment measured into diagonal and
off-diagonal terms (with respect to the quantum number $j$) reveals
interesting physical information about the rotational dynamics. While
$\langle \cos^2\theta\rangle _p$ provides a direct measure of the
rotational population, $\langle \cos^2\theta\rangle _c$ leads to the
temporal evolution of the coherences. By definition,
$\langle\cos^2\theta\rangle _p$ is constant when the pulse is switched
off. In some applications, it can be interesting to maximize only the
permanent alignment.

For this purpose, we use the strategy proposed in
Ref. \cite{sugnydensity}. Considering a sub-Hilbert space
$\mathcal{H}_{j_f}$ of finite dimension defined by the condition
$j\leq j_f$, we introduce the diagonal projection of the
$\cos^2\theta$ operator:
\begin{equation}
\cos^2\theta_p=\sum_{j,m,m'}|j,m\rangle \langle j,m|\cos^2\theta|j,m'\rangle \langle j,m'|,
\end{equation}
such that $\langle\cos^2\theta_p\rangle =\langle\cos^2\theta\rangle _p$.
The target state $\rho_f$ of the control problem is defined as the density matrix maximizing $\langle \cos^2\theta_p\rangle $ and reachable from the initial state $\rho_0$. Due to the constraint of unitary evolution, the density matrices $\rho_0$ and $\rho_f$ have the same spectrum.
In addition, it can be shown that, in this subspace, the two operators $\cos^2\theta_p$ and $\rho_f$ can be simultaneously diagonalized.
One therefore deduces that
\begin{equation}
\textrm{max} [\langle\cos^2\theta_p\rangle ]=\sum_{k=1}^N \chi_k\omega_k,
\end{equation}
where $\chi_1\leq \chi_2\leq \cdots \leq \chi_N$ and $\omega_1\leq
\omega_2\leq \cdots \leq \omega_N$ are the eigenvalues of
$\cos^2\theta_p$ and $\rho_f$, respectively. The integer $N$ is the
dimension of $\mathcal{H}_{j_f}$. If we denote by $|\chi_k\rangle$ the
eigenvectors of $\cos ^2\theta_p$, $\rho_f$ becomes
\begin{equation}
\rho_f=\sum_{k=1}^N \omega_k|\chi_k\rangle \langle \chi_k|.
\end{equation}
Since the subspaces of a given parity of $j$ are not coupled by the
operators $\cos^2\theta_{x,y,z}$, the subdivision
$\mathcal{H}_{j_f}=\mathcal{H}_{j_f}^{(even)}\oplus
\mathcal{H}_{j_f}^{(odd)}$ has to be considered to properly define the
target state, see Ref. \cite{sugnydensity} for details of this construction.

\begin{figure}[tb]
  \centering
  \includegraphics[width=0.98\linewidth]{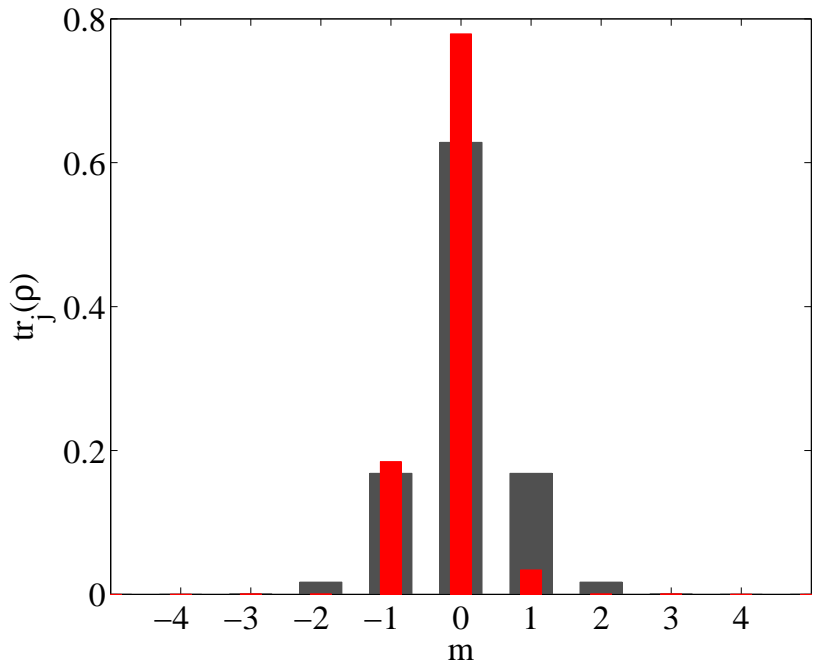}
  \caption{(color online)
    Partial trace $\textrm{Tr}_j[\cdot]$ with respect to the quantum number $j$ of
    initial and target state distributions at $T=5$ K.
    The target distribution corresponds to $\rho_f$, (red or dark gray bar) which
    is the density matrix maximizing the permanent
    alignment at $T=5$ K. The partial trace of the initial density
    matrix is in gray. The parameter $j_f$ is set to 4.}
  \label{fig:themal_distrib}
\end{figure}

Figure \ref{fig:themal_distrib} displays the partial trace $\textrm{Tr}_j[\cdot]$ with
respect to $j$ of the target state $\rho_f$, with $j_f=4$,  for
the CO molecule at $T=5$K. For a given value $m$, this trace is defined by
$\sum_{j=|m|}^{j_{max}}\rho_{jm,jm}^2$. For comparison, we
have also plotted the same distribution for the initial state
$\rho_0$. One clearly sees in Fig. \ref{fig:themal_distrib} that the
optimal distribution is narrowed compared to the thermal
one. Since $\rho_f$ is a diagonal matrix, there is no coherence and we
get $\langle\cos^2\rangle_c =0$. At $T=0$ K, the maximum
permanent alignment is equal to 0.6. This maximum is a
temperature-dependent function, and for $T \neq 0$ K, $ \max[\langle
\cos^2 \theta_p\rangle] \leq 0.6$. For example at $T=5$ K, $
\max[\langle \cos^2 \theta_p\rangle] = 0.518$.

\begin{figure}[tb]
  \centering
  \includegraphics[width=0.98\linewidth]{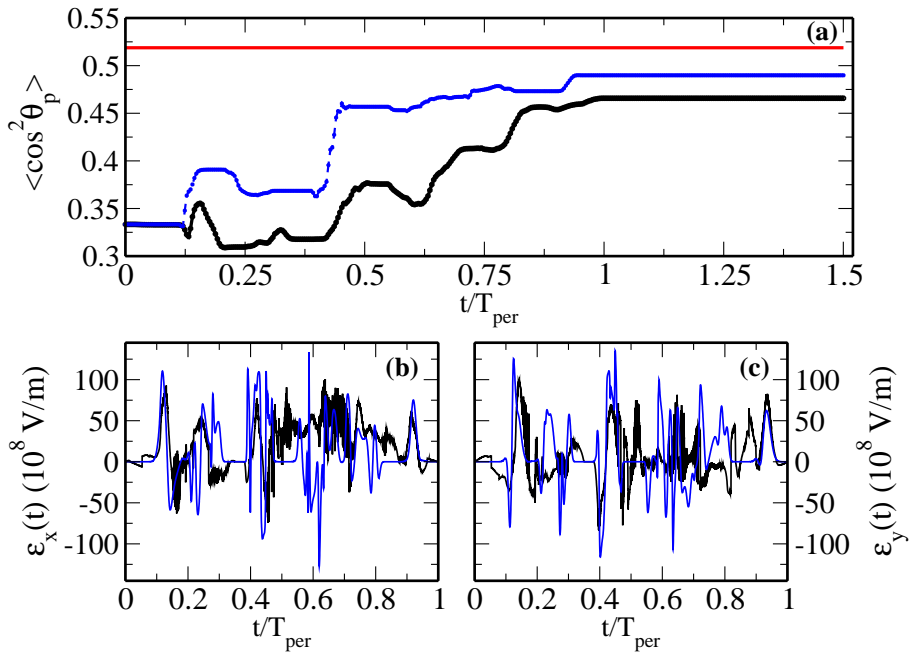}
  \caption{(Color online)
   (a): Permanent alignment dynamics during and after the application of the optimized pulses.
   The black and blue (dark gray) solid lines correspond to the
    dynamics generated by the Lapert and the Krotov optimized pulses. The red (light gray) horizontal solid line depicts here the
    maximum permanent alignment that can be reached in the subspace $\mathcal{H}_{j_f}$. (b): $x$- Components of the optimized fields.
     (c): $y$- Components the optimized fields.
    The black and
    blue (dark gray) curves depict respectively the results derived
    with the Lapert and the Krotov algorithms.}
  \label{fig:dyncos2p}
\end{figure}

We use Eq.~(\ref{eq:hamXY}) and the same guess pulse as in Sec. \ref{subsec:deloc}. The parameter
$j_f$ is fixed to 4.
We have chosen $\lambda$ =  $2\times 10^{-2}$
and   $5\times 10^5$
for the Krotov and Lapert algorithms, respectively.
Figure \ref{fig:dyncos2p} compares
the permanent alignment dynamics for the two algorithms at $T= 5$
K. The pulse is switched off at $t_f=T_{per}$. 
The dynamics are found to be step-like, such that
$\langle\cos^2\theta\rangle _p$ is either constant or varies
suddenly. A permanent alignment of the order of 0.47  and 0.49 is reached for the Lapert
and the Krotov algorithms, respectively for 500 iterations.
In this example,  the comparison of the convergence measured by the final cost is not shown.
We observe as in Fig.~\ref{fig:JT_scheme1} the same behavior for the Lapert
and Krotov algorithms.
As illustrated in Fig. \ref{fig:dyncos2p}, the
optimized solution designed by the Krotov algorithm is simpler than
the one given by the Lapert algorithm.

\section{Summary}
\label{sec:conclusions}
For key control problems of the rotational dynamics, we have designed in this work different optimized pulses from two monotonically convergent algorithms, the Lapert and the Krotov ones. Our numerical findings confirmed by the three examples discussed in this work are as follows:
\begin{enumerate}
\item The final fidelities reached by the two algorithms are very similar.
\item The Lapert algorithm is somehow more flexible compared to the
  Krotov one in the sense that the parameter $\lambda$ can be chosen
  without any constraints, while this parameter has to satisfy the
  relation (\ref{eq:cond-delta2}) in the Krotov approach in order to
  ensure monotonic convergence.
\item The Krotov algorithm is more efficient than the Lapert one in terms of CPU time.
\item The optimized Krotov field has a simpler structure than the Lapert one, which generally presents some unwanted oscillatory behaviors.
\end{enumerate}
Having in mind the work of Ref. \cite{manai}, an open question is now the application of these optimal control algorithms to more complex systems. The computation of the optimal field allowing the cooling of rovibrational dynamics could be an interesting test case, in particular because non unitary processes have to be taken into account.\\

\noindent \textbf{Acknowledgment}\\
  The authors thank Daniel Reich for many helpful
  discussions and his comments on the manuscript.
  Financial supports from the Conseil R\'egional de Bourgogne and
  the QUAINT coordination action (EC FET-Open) are gratefully acknowledged.

\appendix
\section{Description of monotonically convergent
  algorithms}\label{app}
The Krotov and Lapert optimization algorithms are summarized here
for pure state quantum dynamics. This description is straightforwardly
extended to the density matrix formalism. To simplify notation,
we restrict ourself to the maximization of the projection onto a
target state.
The two algorithms can analogously be used for maximization or
minimization  of the expectation value of a given observable.
The control problem is characterized by  maximization of
the functional $C$,
\begin{equation}
    \label{eq:J}
      C =
      C_{t_f} \left [\{|\psi(t_f)\rangle\}\right]  
      - \int_0^{t_f}   C_t\left[ \{|\psi(t)\rangle\}, E(t)\right]\;dt,
\end{equation}
where $C_{t_f}$ is the final time cost functional and
$C_t$ the running cost. The parameter
$E$ denotes the external field and
$|\psi(t)\rangle$ the wave function describing the state of the system
at time $t$. Its time evolution is governed by Eq.~(\ref{eq:evolwp}).
If $|\psi_f\rangle$ is the  target state, the final cost can be defined as follows:
\begin{equation}
    \label{eq:JT}
    C_{t_f} \left [\{|\psi(t_f)\rangle\}\right] =
          \left | \langle \psi_f |\psi(t_f)\rangle\right|^2.
 \end{equation}
Here, only a running cost which does not depend on the state of the
system is considered:
$$
C_t\left[ \{|\psi(t)\rangle\}, E(t)\right] = g\left[
E(t)\right].
$$
Extension to a state-dependent running cost is
described in Ref \cite{reich}.
The main difference between the two algorithms is in
 the choice of the running cost.

\subsection{Krotov's method}
\label{subsec:krotov}

The derivation of the Krotov algorithm presented here follows closely
Ref \cite{reich}, specializing it to
a non-linear interaction of the system with the control field, see
Eqs.~(\ref{eq:ham1}) and (\ref{eq:hamXY}). There is no requirement for
a specific power of $E(t)$
the running cost $g$ so we choose it to minimize the change in the
energy of the field~\cite{JosePRA03},
\begin{equation}
   \label{eq:ga-kr}
   g\left[E(t)\right] =  \frac{\lambda}{S(t)}\left(E(t)-E_{ref}\right)^2,
\end{equation}
with $E_{ref}$ denoting a reference field,  $S(t)$ a shape function
and $\lambda$ a weight.

Krotov's method is based on the construction of an auxiliary functional,
$\mathcal L\left[\{|\psi\rangle\}, E, \Phi\right]$ with
$\Phi$ an arbitrary functional. It is chosen such that the
maximization of $\mathcal{L}$ is equivalent to
maximization of $C$ of Eq~(\ref{eq:J}). $\Phi$ is used to ensure a
global minimum with respect to changes in the state. Then any
change in the state will lead to an increase in the value of
$\mathcal{L}$, i.e., to monotonic
convergence~\cite{KonnovAC99,SklarzPRA02}.
This is achieved by expanding $\Phi$ to second
order in the change of the state, $|\Delta\psi(t)\rangle$,
\begin{eqnarray}
  \label{eq:Phi}
\Phi\left[\{|\psi\rangle\},t\right] &=&
   \langle \chi(t)|\psi(t)\rangle
    + \langle\psi(t)|\chi(t)\rangle \nonumber \\
    && +\frac{1}{2} \langle\Delta\psi(t)|\bold{\sigma}(t)|
  \Delta\psi(t)\rangle  \,.
\end{eqnarray}
When the equations of motion are linear with respect to the state and the running
cost functional does not depend on the state, the second order contribution is not required~\cite{reich}.
Since, in this work,  we consider a  quantum control problem which fulfills these conditions, a
first order construction of $\Phi$ is sufficient.
The construction of $\Phi$ is described in detail in Ref. \cite{reich}.
The auxiliary functional is defined by:
\begin{eqnarray}
  \mathcal L[\{|\psi\rangle\},E,\Phi] &=&
  G\left[\{|\psi(t_f)\rangle\}\right] 
-\Phi\left[\{|\psi(0)\rangle\},0\right]  \nonumber \\
&& - \int_0^{t_f}  R\left[\{|\psi(t)\rangle\},E(t),t\right]dt,
\label{eq:defL}
\end{eqnarray}
where the final time functional and running functional, $G$ and $R$,
are given by
\begin{equation}
   \label{eq:G}
   G(\{|\psi(t_f)\rangle\}) =
   C_{t_f} \left [\{|\psi(t_f)\rangle\}\right]
         +\Phi\left[\{|\psi(t_f)\rangle\},t_f\right]\,,
\end{equation}
\begin{eqnarray}
   \label{eq:R}
   R\left[\{|\psi(t)\rangle\},E(t),t\right] &=&
   \frac{\partial\Phi}{\partial t} + g\left[E(t)\right] \nonumber
   -i \left(\nabla_{|\psi\rangle}\Phi\right) |H \psi(t)\rangle
   \nonumber \\
   &&+ i \langle \psi(t)| H \left(\nabla_{\langle\psi|}\Phi\right)
\end{eqnarray}
For a maximization problem, the following conditions have to be fulfilled:
\begin{equation}
  \label{eq:minL}
  \mathcal L[\{|\psi_{k}\rangle\},E_{k},\Phi] \le
  \mathcal L[\{|\psi_{k+1}\rangle\},E_{k+1},\Phi],
\end{equation}
where $k$ indicates the iterative step.
Sufficient conditions for maximizing $\mathcal L$ translate into
maximizing $G$ and minimizing $R$ at each time:
\begin{eqnarray}
  \label{eq:Delta}
  \mathcal L[\{|\psi_{k+1}\rangle\},E_{k},\Phi]-
  \mathcal L[\{|\psi_{k}\rangle\},E_{k+1},\Phi]
  &=&
  \nonumber \\
  \Delta_1+
  \int_0^{t_f} \Delta_2(t)dt +
  \int_0^{t_f} \Delta_3(t)dt,
\end{eqnarray}
where the $\Delta_i$, $i=1,\cdots,3$ are given by:
\begin{eqnarray}
  \label{eq:Delta1}
\Delta_1 =  G\left[\{|\psi_{k+1}(t_f)\rangle\}\right]-
G\left[\{|\psi_{k}(t_f)\rangle\}\right],
\end{eqnarray}
\begin{eqnarray}
  \label{eq:Delta2}
\Delta_2(t) &=&   R\left[\{|\psi_{k+1}(t)\rangle\},E_{k}(t),t\right]
  \nonumber \\
  && - R\left[\{|\psi_{k+1}(t)\rangle\},E_{k+1}(t),t\right],
\end{eqnarray}
and
\begin{eqnarray}
  \label{eq:Delta3}
  \Delta_3(t) &=& R\left[\{|\psi_{k}(t)\rangle\},E_{k}(t),t\right]
  \nonumber \\
&&- R\left[\{|\psi_{k}(t)\rangle\},E_{k+1}(t),t\right].
\end{eqnarray}
Non-negativeness of the $\Delta_i$ ($i=1,\cdots,3$) ensures
monotonic convergence. For a quantum control problem where the cost functional is state-independent and the equations of motion are linear with respect to the state, positivity
of $\Delta_1$ and $\Delta_3$ is automatically satisfied \cite{SklarzPRA02, JosePRA03}.
Non-negativeness of $\Delta_2$ can be obtained by a proper choice of
$\lambda$ and the shape function $S(t)$ \cite{reich}:
\begin{equation}
   \label{eq:cond-delta2}
  \frac{\lambda}{S(t)} > \Delta_{M_2^E},
\end{equation}
where $\Delta_{M_2^E}$ is the spectral radius of ${M_2^E}= \partial^2
H\left[E(t)\right]/\partial E^2$.

Evaluating the extremum condition for $\mathcal L$ yields the control equations,
\begin{enumerate}
\item Equation of the control field:
  \begin{eqnarray}
    \label{eq:fieldnew}
    \frac{\partial g}{\partial E}\bigg|_{E_{k+1}}(t) &&= \\\nonumber
    && 2\mathfrak{Im}
    \left[\left\langle \chi_{k}(t)\left|
        \frac{\partial H\left[E(t)\right]}{\partial E}\bigg|_{E_{k+1}}
      \right|\psi_{k+1}(t)\right\rangle
       \right].
   \end{eqnarray}
   \item Equation of motion for the adjoint state $|\chi\rangle$, with
     'initial'      condition:
       
     \begin{eqnarray}\label{eq:evolchi}
       i\frac{\partial}{\partial t} |\chi_k(t)\rangle
       &=& H\left[ E_{k}(t)\right]|\chi_{k}(t)\rangle\, \\
      \qquad |\chi_{k}(t=T)\rangle &=& -\nabla_{\langle\psi_{k}|}
      C_{t_f}\big|_{|\psi_{k}\rangle} .
    \end{eqnarray}
 \item Equation of motion of the  state $|\psi\rangle$, with  initial
   condition $|\psi_{ini}\rangle$
       \begin{eqnarray}\label{eq:evolpsi}
       i\frac{\partial}{\partial t} |\psi_{k+1}(t)\rangle
       &=& H\left[ E_{k+1}(t)\right]|\psi_{k+1}(t)\rangle\, \\
      \qquad |\psi_{k+1}(t=0)\rangle &=& |\psi_{ini}\rangle .
    \end{eqnarray}
    \end{enumerate}
At each iteration, the update of the field $E_{k+1}(t)$ obtained
from Eq.~(\ref{eq:fieldnew}) involves a backward
and a forward propagation, Eq.~(\ref{eq:evolchi}) and Eq.~(\ref{eq:evolpsi}),
respectively. If the system interacts non-linearly with the control field,
both left and right hand sides of Eq.~(\ref{eq:fieldnew}) depend on $E_{k+1}(t)$.
For simplicity, we assume that the change of the control field
between iterations $k$ and $k+1$ is small enough such that
$\partial H(E(t))/\partial E|_{E_{k+1}} \approx \partial
H(E(t))/\partial E|_{E_{k}}$.
\subsection{The Lapert approach}
\label{subsec:lapert}
\subsubsection{General description}
 While in the
Krotov method, this running cost minimizes the change in the energy of the field, in the
Lapert algorithm, this choice is different.
Basically, $g$ is chosen so as Eq.~(\ref{eq:fieldnew}) admits a real
solution at any time $t$.
The cost is defined as follows:
\begin{equation}
   \label{eq:ga-lp}
   g\left[E(t)\right] =  \frac{\lambda}{S(t)}\left(E(t)-E_{ref}\right)^{2n}\,,
\end{equation}
For a Hamiltonian given by Eq.~(\ref{eq:ham1}), Eq.~(\ref{eq:fieldnew}) leads to
\begin{eqnarray}
  \label{eq:newfield-poly}
  &&2n \frac{\lambda}{S(t)}\left(E_{k+1}(t)-E_{k}(t)\right)^{2n-1}-
  \nonumber \\
  && 2\Im
       \left[\langle \chi_{k}|\bold{\mu}+2\bold{\alpha}E_{k+1} + 3 \bold{\beta}E_{k+1}^2|\psi_{k+1}\rangle
       \right] = 0
\end{eqnarray}
The left hand side of  Eq.~(\ref{eq:newfield-poly}) can be viewed as
a polynomial of $E_{k+1}$. Choosing the integer $n$ such that
$\lambda/S(t)E_{k+1}^{2n-1}$ is a monomial of order higher than the
right hand side of  Eq.~(\ref{eq:newfield-poly}) ensures that there
exists a real solution to the equation at each time $t$. For a
non-linearity of order 3, $n=2$ is sufficient. The
 conditions for monotonic convergence are determined
through variation of  $\Delta C$ given by:
\begin{eqnarray}
  \label{eq:DeltaJ}
  \Delta C &=& C_{k+1}-C_k
  \nonumber \\
  &=&
  C_{t_f}\left [\{|\psi_{k+1}(t_f)\rangle\}\right]-
  C_{t_f}\left [\{|\psi_{k}(t_f)\rangle\}\right]
  \nonumber \\
  && +
   \int_0^{t_f}   g\left[E_{k+1}(t)\right]-
  g\left[E_{k}(t)\right]\; dt.
\end{eqnarray}
which needs to be positive \cite{lapertalgo}.

\subsubsection{Role of the parameter $\lambda$}
\begin{figure}[tb]
  \centering
  \includegraphics[width=0.98\linewidth]{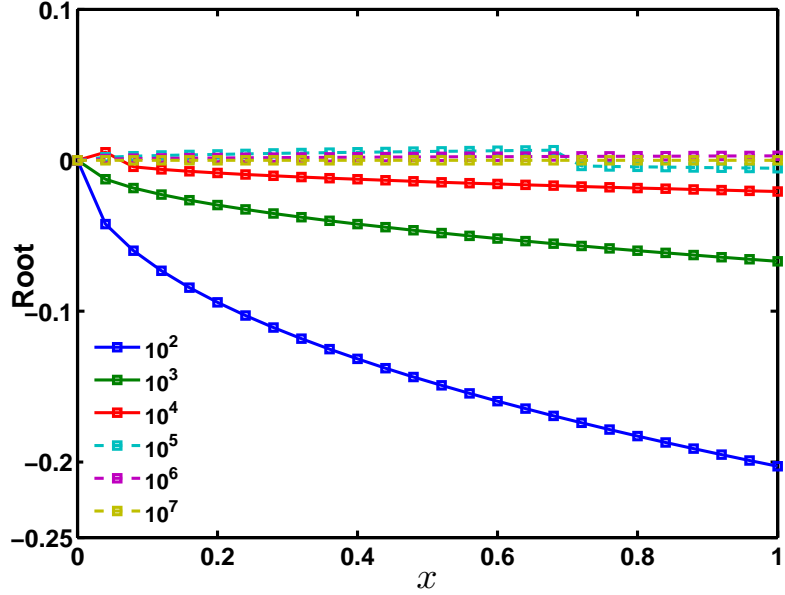}
  \caption{(Color online)
    Variation of one of the real roots of Eq.~(\ref{eq:rootsanalysis}) as
    a function of  $x=\langle\chi(t)|\psi(t)\rangle$ for different values of
    $\lambda$, $10^2\le \lambda \le 10^7$.}
  \label{fig:variationroots}
\end{figure}

We discuss in this section the way the parameter $\lambda$ affects the
optimized solution. For this purpose, we have analyzed the behavior of
one of the real roots of  the polynomial
Eq.~(\ref{eq:newfield-poly}). To simplify the description, the
operators $\vec{\mu}$, $\bold{\alpha}$ and $\bold{\beta}$ have been
replaced by their maximum eigenvalues. For $n=2$,
Eq.~(\ref{eq:newfield-poly}) can then be written as follows:
\begin{eqnarray}
  \label{eq:rootsanalysis}
  &&4 \frac{\lambda}{S(t)}\left(E_{k+1}(t)-E_{k}(t)\right)^{3}-
  6 |\beta|_{\mathrm{max}} \Im[\langle \chi_{k}|\psi_{k+1}\rangle] E_{k+1}^2
  \nonumber \\
  &&
  -4 |\alpha|_{\mathrm{max}} \Im[\langle \chi_{k}|\psi_{k+1}\rangle] E_{k+1}
  \nonumber \\
  &&
  -2|\mu|_{\mathrm{max}} \Im[\langle \chi_{k}|\psi_{k+1}\rangle]
        = 0\,.
\end{eqnarray}
 Figure \ref{fig:variationroots} illustrates the
variation of one of the real roots of Eq.~(\ref{eq:rootsanalysis}) as
a function of  $x=\langle\chi(t)|\psi(t)\rangle$ for different values of
$\lambda$.
The range of $\lambda$ is taken from $10^2$ to $10^7$. For
large values of $\lambda$, the variation of the root is very slow with
respect to $x$ while for a
value smaller then $10^4$, the change of the roots can be very fast. This observation qualitatively explains the fast oscillations occurring in the Lapert optimized fields.


\end{document}